\documentclass[namedreferences]{SolarPhysics}
\usepackage[optionalrh]{spr-sola-addons} % For Solar Physics
\usepackage{graphicx}
\usepackage{color}           % For color text: \color command
\usepackage{url}             % For breaking URLs easily trough lines
\newcommand{\aap}{    {\it Astron. Astrophys.}}

\newcommand{\apj}{    {\it Astrophys. J.}}
\newcommand{\apjl}{   {\it Astrophys. J. Lett.}}

\newcommand{\grl}{    {\it Geophys. Res. Lett.}}

\newcommand{\jgr}{    {\it J. Geophys. Res.}}

\newcommand{\solphys}{{\it Solar Phys.}}

\newcommand{\ssr}{    {\it Space Sci. Rev.}}

\begin{document}

\begin{article}

\begin{opening}

\title{Relationship between the Magnetic Flux of Solar Eruptions
and the Ap Index of Geomagnetic Storms}

\author{I.M.~\surname{Chertok}$^{1}$\sep
    M.A.~\surname{Abunina}$^{1}$\sep
    A.A.~\surname{Abunin}$^{1}$\sep
    A.V.~\surname{Belov}$^{1}$\sep
    V.V.~\surname{Grechnev}$^{2}$
        }

\runningauthor{Chertok et al.}
 \runningtitle{Solar eruptions and the Ap geomagnetic index}

\institute{${}^{1}$Pushkov Institute of Terrestrial Magnetism,
            Ionosphere and Radio Wave Propagation (IZMIRAN), Troitsk, Moscow
            Region, 142190 Russia email: \url{ichertok@izmiran.ru}\\
            ${}^{2}$Institute of Solar-Terrestrial Physics SB RAS,
            Lermontov St.\ 126A, Irkutsk 664033, Russia email: \url{grechnev@iszf.irk.ru}
}

\begin{abstract}
Solar coronal mass ejections (CMEs) are main drivers of the most
powerful non-recurrent geomagnetic storms. In the
extreme-ultraviolet range, CMEs are accompanied by bright
post-eruption arcades and dark dimmings. The analysis of events of
the Solar Cycle 23 (Chertok \textit{et al.}, 2013, \textit{Solar Phys}.
\textbf{282}, 175) revealed that the summarized unsigned
magnetic flux in the arcades and dimming regions at the
photospheric level, $\Phi$, is significantly related to the
intensity (Dst index) of geomagnetic storms. This provides the
basis for the earliest diagnosis of geoefficiency of solar
eruptions. In the present article, using the same data set, we
find that a noticeable correlation exists also between the
eruptive magnetic flux, $\Phi$, and another geomagnetic index, Ap.
As the magnetic flux increases from tens to $\approx 500$ (in
units of $10^{20}$ Mx), the geomagnetic storm intensity measured
by the 3-hour Ap index, enhances in average from Ap $\approx 50$
to a formally maximum value of 400 (in units of 2 nT). The
established relationship shows that in fact the real value of the
Ap index is not limited and during the most severe magnetic storms
may significantly exceed 400.
\end{abstract}


\end{opening}

%%%%%%%%%%%%%%%%%%%%%%%%%%%%%%%%%%%%%%%%%%%%%%%%%%%%%%%%%%%%%%%%%%%%%%%%%%%%%%

\section{Introduction}
 \label{S-introduction}

Intense non-recurrent geomagnetic storms (GMSs) are the most
significant space weather disturbances. In contrast to relatively
weak recurrent storms associated with high-speed solar wind
streams from coronal holes, they are initiated by coronal mass
ejections (CMEs) and their interplanetary extensions (ICMEs)
(\textit{e.g.}, \opencite{BothmerZhukov2007};
\opencite{Gopalswamy2009}). A GMS occurs when a large and fast
CME/ICME brings to the Earth the magnetic field with a
sufficiently strong and prolonged southward (negative) $B_z$
component. One of the most important tasks of the
solar-terrestrial physics and space weather prediction is
diagnostics of geoeffectiveness of CMEs, \textit{i.e.},
quantitative forecast of a possible non-recurrent GMS from
observed characteristics of the eruption that just occurred.
Existing algorithms of such diagnostics are based in one way or
another on the measurements of the CME speed and shape in the
plane of the sky near the Sun \cite{Gopalswamy2009, Kim2010}.

We develop another approach to the early diagnostics of solar
eruptions, in which quantitative characteristics of such
large-scale CME manifestations as post-eruption (PE) arcades and
dimmings observed in the extreme ultraviolet (EUV) range are used
as key parameters instead of the projected CME speed and shape
\cite{ChertokGrechnev2006,Ch2013}. Dimmings are CME-associated
regions in which the EUV (and soft X-ray) brightness of coronal
structures is temporarily reduced during an ejection and persists
over many hours \cite{Thompson1998, Webb2000, HudsonCliver2001}.
The deepest stationary long-lived dimmings adjacent to the
eruption center are interpreted mainly as a result of plasma
outflow and density decrease in footpoints of erupting and
expanding CME flux ropes. Large-scale arcades of bright loops
enlarging in size over time arise at the place of the main body of
pre-eruption magnetic flux ropes ejected as CMEs \cite{Kahler1977,
Sterling2000, Tripathi2004}. As a whole, PE arcades and dimmings
visualize the structures and areas involved in the processes of
CME eruptions.

A number of studies indicate that parameters of CMEs and
near-Earth magnetic clouds (MCs) are governed by total magnetic
fluxes in their solar source regions.
\inlinecite{QiuYurchyshyn2005} found a correlation between the CME
speed and the total reconnection flux in associated flares.
\inlinecite{Leamon2004} and \inlinecite{Mandrini2005} demonstrated
that the total magnetic fluxes in MCs estimated from \textit{in
situ} measurements near the Earth and their solar sources were
close to each other. \inlinecite{Webb2000} and
\inlinecite{Attrill2006} found a similarity between the magnetic
fluxes in MCs and dimming regions. \inlinecite{Qiu2007} and
\inlinecite{Hu2014} found this similarity for the magnetic
reconnection flux in several events. Therefore, a correspondence
is expected between the magnetic flux involved in an eruption, on
the one hand, and parameters of the associated geospace
disturbances, on the other hand. If a magnetic cloud carries a
negative $B_z$ component, then a correlation might be expected
between the magnetic fluxes in eruption regions and the
intensities of the GMSs that they produce.

\inlinecite{Ch2013} established that a statistical dependence of
the GMS intensity measured by the Dst index (as well as of the
onset and peak GMS transit times and Forbush decrease  magnitudes)
on the eruptive magnetic flux in the arcades and dimming regions,
$\Phi$, exists indeed. The aim of the present work is to analyze
the relationship between this eruptive flux $\Phi$ and another
geomagnetic index Ap that also characterizes the GMS intensity and
is often used in the solar-terrestrial forecasting.

\section{Analyzed Parameters and Data}

Recall that the hourly storm-time disturbance index Dst (see
\url{http://wdc.kugi.kyoto-u.ac.jp/dstdir/index.html}) is
calculated from data of four low-latitude geomagnetic
observatories and characterizes the effect of the global
equatorial ring current. The latter is manifested in the inner
magnetosphere and causes a decrease of the horizontal component of
the terrestrial magnetic field during the main phase of GMSs. The
linear planetary 3-hour Ap index, considered below, characterizes
the strength of auroral currents. The Ap index  is defined as the
mean value of the variations of the terrestrial magnetic field,
which corresponds to a logarithmic Kp index, measured by data of
13 geomagnetic stations located at moderately high geomagnetic
latitudes mainly in the northern hemisphere \cite{Siebert1996,
Zabol}. The Ap index varies from 0 to a formally maximum value of
400 corresponding to Kp = 9, and is measured in units of 2 nT
(hereinafter, as is often done, we will omit the units). We
emphasize that the formal upper limit of Ap = 400 is is a
condition that appears as a result of the conversion of Ap from Kp
indices. In a physical sense the Ap index is not limited. It is
determined by real variations of the terrestrial magnetic field
components and during the strongest magnetic storms can
considerably exceed 400. Features of the Dst and Ap indices are
such that relationships between them in concrete events may be
different, depending on the impact of an ICME on the equatorial
and auroral current systems. In addition, the peak time of the Dst
and Ap indices are not always identical and may differ by a few
hours.

We consider the same ensemble of the events that has been analyzed
by \inlinecite{Ch2013}: strong non-recurrent GMSs during Solar Cycle 23
(1996--2008) with Dst~$\leq -100$~nT, which
reliably or with a high probability are identified with a concrete
eruption from the central zone of the visible solar hemisphere
within $\pm 45^{\circ}$ from the central meridian. We will
distinguish events caused by eruptions that occurred in active
regions (ARs), and events associated with filament eruptions outside
ARs referring to them as ``AR events'' and ``non-AR events'',
respectively. These two categories of events differ significantly in
the characteristics of accompanying arcades and dimmings, properties
of CMEs/ICMEs, and intensity of GMSs that they cause (\textit{e.g.},
\opencite{Svestka2001}).

Our source data are the EUV solar images obtained with the
\textit{Extreme ultraviolet Imaging Telescope} (EIT:
\opencite{Delab1995}) in the 195~\AA\ channel and the magnetograms
obtained with the \textit{Michelson Doppler Imager} (MDI:
\opencite{Scherrer1995}), both on the \textit{Solar and
Heliospheric Observatory} (SOHO: \opencite{Domin1995}). For each
of the analyzed events, after a routine processing of
corresponding FITS files, the solar rotation in the analyzed
images was compensated, and then the same fixed image before an
eruption was subtracted from all subsequent ones to obtain
fixed-base difference images \cite{ChertokGrechnev2005}.
Significant regions of the arcades and dimmings are extracted
following selected relative (not absolute) criteria of
brightness variations. The analysis showed that a brightness
depression of more than 40\% from the pre-eruption level was an
optimal criterion for extraction of relevant significant dimmings.
For PE arcades, an appropriate criterion was that which
extracted the area around the eruption center where the brightness
in the 195~\AA\ channel exceeded 5\% of the maximum one. A total
(unsigned) eruptive magnetic flux within the extracted arcade and
dimming areas was evaluated at the photospheric level by
co-aligning the resulting EIT difference images with an MDI
line-of-sight magnetogram recorded just before an eruption. The
detailed description of all required procedures is given in
\inlinecite{Ch2013}.

In this case we will characterize the maximum intensity of GMSs by
a peak in the 3-hour Ap index, the values of which are determined
in the GeoForschungsZentrum (GFZ), Potsdam, and presented at
\url{ftp://ftp.gfz-potsdam.de/pub/home/obs/kp-ap/wdc/}. The
comprehensive list of the analyzed events and their parameters are
shown in a table displayed at the site
\url{http://www.izmiran.ru/~ichertok/Ap/}. The table includes 90
events and contains in particular the following data: the numbers
of the GMS events according to the Coordinated Data Analysis
Workshop (CDAW) catalog \cite{Zhang2007}; information on date,
time, and intensity of a GMS, including Ap index; data about the
eruption, which reliably or with a high probability was a source
of the given GMS; the total magnetic fluxes in the arcades and
dimming regions; and some others.

\section{Results}

\begin{figure} % {15}
  \centerline{\includegraphics[width=\textwidth]
   {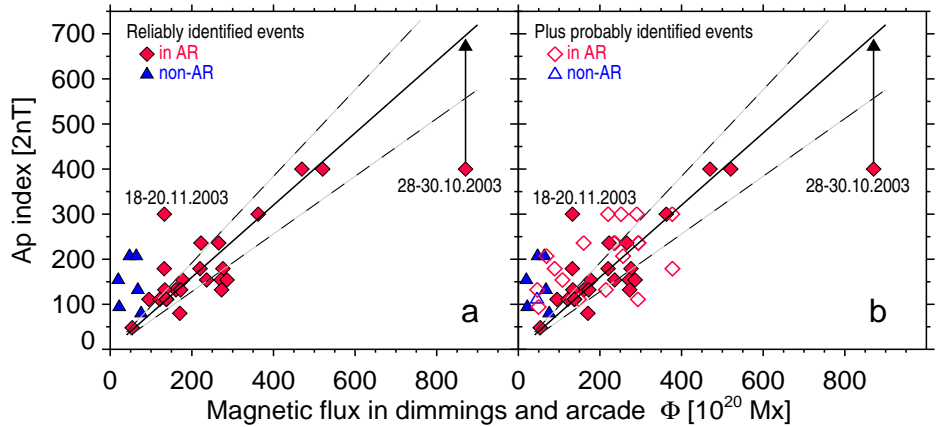}
  }
  \caption{Dependence of the geomagnetic Ap index on the total
magnetic flux in the arcades and dimming regions $\Phi$: (a) for
GMSs reliably identified with a definite solar eruption (filled
symbols) and (b) for all considered events including those with
probable solar source identification (open symbols). Red diamonds
denote eruptions in ARs, and blue triangles denote eruptions of
filaments outside ARs. The dashed lines delimit the adopted 20\%
deviation band.}
  \label{F-fig1}
  \end{figure}

First, we consider the events, in which a GMS is reliably
identified with a particular solar eruption. Figure~\ref{F-fig1}a
shows the location of such events on the plot of the geomagnetic
Ap index \textit{vs.} the eruptive magnetic flux in the arcades
and dimming regions, $\Phi$''.  In this figure and the following
ones, the diamonds denote the AR eruptions, and the triangles
correspond to the non-AR eruptions.

%%%%%%%%%%%%%%%%%%%%%%%%%%%%%%%%%%

As noted by \inlinecite{Ch2013}, few events associated with
filament eruptions outside ARs (non-AR events) form a separate
group, which is markedly different from the events produced by the
eruptions in ARs (AR events). The non-AR events are characterized
by relatively low magnetic fluxes, $\Phi < 80 \times 10^{20}$~Mx,
but they can cause GMSs with Ap~$> 200$. Figure~\ref{F-fig1} does
not show any pronounced dependence of Ap on $\Phi$ for such
events. A possible reason for these features of the non-AR
events is that the criteria for the extraction of significant
arcades and dimmings accepted for the AR events are not entirely
suitable for the non-AR events. It is known \cite{Svestka2001}
that filament eruptions outside ARs are accompanied by relatively
weak, but more extensive PE arcades, as well as by shallower
dimmings in comparison with eruptions in ARs. This suggests that
the relative thresholds of the brightness variations for the
extraction of the arcades and dimmings indicated above should be
optimized at lower levels for the non-AR events. In this case, the
values of the eruptive magnetic flux would noticeably increase,
but in different ways for different events. Some other factors may
also affect the characteristics of GMSs caused by the non-AR
eruptions. Therefore, the non-AR events require a special
additional analysis and will not be counted below in this paper.

%%%%%%%%%%%%%%%%%%%%%%%%%%%%%%%%%%

Among the AR events (diamonds), attention should be paid to an
event that strongly deviates from other GMSs.  This is a famous
exceptional event of 18--20 November 2013, in which under
relatively small eruptive magnetic flux $\Phi \approx 133 \times
10^{20}$ Mx the GMS intensity reached Ap $\approx 300$ (this storm
was the strongest in Solar Cycle 23 in terms of the index Dst
$\approx -422$~nT). The causes of the exceptional properties of
this extreme event are analyzed in detail by
\inlinecite{Grechnev2014}. They, in particular, showed that the
GMS was caused by arrival to the Earth of a compact spheromak with
a magnetic field strength up to $|B| \approx 56$~nT and a large,
prolonged south component $B_z \approx -46$ nT. Such a strong
field in the interplanetary magnetic cloud was preserved due to
its anomalously weak expansion in the propagation from the Sun to
the Earth. Below we do not consider this event because of its
atypical properties.

The remaining reliably identified AR events, shown in
Figure~\ref{F-fig1}a, demonstrate a significant relation between
the eruptive flux $\Phi$ and the geomagnetic index Ap. When the
flux $\Phi$ increases from tens to about 500 (in $10^{20}$~Mx
units), the Ap index rises up to the formally maximum value of
400. Three events with the largest eruptive flux indicate this
dependence at the formal peak level Ap = 400. Real values of the
Ap index for these three events are not given in catalogs. We
discard temporarily the 28--30 October 2003 event with the largest
magnetic flux $\Phi$ and assume that for two events with the
magnetic flux $\Phi$ in a range of $(470-520) \times 10^{20}$~Mx
the formal index Ap = 400 was close to the real one. Then the
dependence of the Ap index on the eruptive magnetic flux $\Phi$
for the AR events can be expressed by a linear relation
Ap~$= 0.8\Phi$ (here again $\Phi$ is expressed in units of
$10^{20}$~Mx). In this case, the correlation coefficient between
the observed and calculated values of Ap reaches $r \approx 0.90$.
For additional evaluation of the scatter in data points, we choose
a deviation band bounded by $\pm 0.2$ from the regression line
that appears to be acceptable for the forecasting. Calculations
show that 14 out of 22 events (\textit{i.e.}, 64\%) fall into this
deviation band. 

Extrapolation of the obtained relation allows us to estimate a
probable increase of the real Ap index for eruptions with the
largest magnetic flux without the formal upper limit of Ap = 400.
In particular, for the 28--30 October 2003 event with $\Phi
\approx 870 \times 10^{20}$~Mx, the Ap index, in accordance with
this dependence, should be around 700. This estimate does not
contradict the largest variations of the horizontal component of
the Earth's magnetic field (about 1700--2500 nT) registered by
high-latitude magnetometers (\textit{e.g.},
\opencite{Panasyuk2004}). According to the data of the magnetic
observatory Moscow (IZMIRAN), in the period 21--24~UT on 29
October 2003 the field variations were about 1200 nT, that yields
Ap~$\approx 800$. This also agrees well with the dependence
presented above.

The dependence of the geomagnetic Ap index on the eruptive
magnetic flux $\Phi$ for the AR events appears to be basically the
same, when the GMSs with a probable source identification (open
symbols in Figure~\ref{F-fig1}b) are added to the unambiguously
identified ones. Here, as expected, the scatter of the points
increases, and the correlation coefficient between $\Phi$ and Ap
reduces to $r \approx 0.73$. In this case, 19 points out of 41
(\textit{i.e.}, 46\%) fall into the same deviation band.

\section{Conclusion}

Our analysis has shown that for AR events a noticeable statistical
relationship exists between the GMS intensity measured by the Ap
index and the eruptive magnetic flux $\Phi$ in solar EUV arcades
and dimming regions. This relationship also demonstrates that
solar eruptions with the greatest magnetic flux can lead to the
strongest GMSs, in which a real Ap index significantly exceeds the
formally introduced maximum level of Ap = 400.

This result is consistent with those of the paper by
\inlinecite{Ch2013}, in which the dependencies were established on
the same eruptive magnetic flux, $\Phi$, for such important space
weather parameters as: i)~the intensity of GMSs characterized by
another geomagnetic index Dst, ii)~the values of the Forbush
decreases (FDs) of the galactic cosmic ray flux, and iii)~the
onset and peak times of these disturbances. This means that, in
spite of many factors affecting the propagation of interplanetary
clouds from the Sun to the Earth and the character of their
interaction with the Earth's magnetosphere, the parameters of
non-recurrent GMSs (and FDs) caused by CMEs/ICMEs are largely
determined by the power of solar eruptions (in terms of the total 
magnetic flux in dimmings and arcades). This is especially
conspicuous for sufficiently large eruptions. Just due to this
fact, the statistical relationships exist between the magnetic
flux in the source region of a CME, on the one hand, and the main
parameters of GMSs and FDs, on the other hand.

The positive results of the present work and the analysis of
\inlinecite{Ch2013} demonstrate the relevance of the of the
magnetic flux in arcades and dimmings as a diagnostic parameter.
This parameter can be used for the earliest estimations of the
intensity and temporal characteristics of GMSs (and FDs),
including both Dst and Ap indices, even without taking into
account the information on associated CMEs and factors determining
the $B_z$ component. At the present stage, we have considered only
sufficiently intense GMSs. Therefore, when the GMS amplitude is
estimated on the basis of the dependencies established in this article 
and by \inlinecite{Ch2013}, it refers to a value close to the maximum
that can be expected, provided that an ICME contains a significant
negative (south) $B_z$ component. The proposed tool for the early
diagnostics of the geoeffectiveness of a solar eruption should be
considered and used as an initial component of a wider complex of
methods of the short-term GMS and FD forecasting including also
those based on the measurements of near-the-Sun CMEs, MHD models,
stereoscopic observations of ICME propagation, and others.

\begin{acks}
We are grateful to an anonymous reviewer for constructive
comments, which helped us to improve the manuscript. The authors
thank the SOHO EIT and MDI teams as well as the CDAW participants
for data and materials used in the present study. SOHO is a project
of international cooperation between ESA and NASA. This research was
supported by the Russian Foundation of Basic Research under grants
12-02-00037, 14-02-00367, and the Ministry of education and science
of Russian Federation under projects 8407 and 14.518.11.7047.
\end{acks}

\end{article}

\end{document}